\documentclass[%
 reprint,
 amsmath,amssymb,
 aps,
]{revtex4-2}

\usepackage{graphicx}
\usepackage{dcolumn}
\usepackage{bm}

\begin{document}


\title{Precision test of the weak interaction with slow muons} 

\author{ Xin Chen }
\email{xin.chen@cern.ch}
\altaffiliation[also at ]{Center for High Energy Physics, Tsinghua University, Beijing 100084, China}
\affiliation{Department of Physics, Tsinghua University, Beijing 100084, China}
\author{ Zhen Hu }
\email{zhen.hu@cern.ch}
\altaffiliation[also at ]{Center for High Energy Physics, Tsinghua University, Beijing 100084, China}
\affiliation{Department of Physics, Tsinghua University, Beijing 100084, China}
\author{ Yongcheng Wu }
\email{ycwu@njnu.edu.cn}
\affiliation{ Department of Physics and Institute of Theoretical Physics, Nanjing Normal University, Nanjing 210023, China }
\author{ Hui Li }
\affiliation{Department of Physics, Tsinghua University, Beijing 100084, China}
\author{ Shaogang Peng }
\affiliation{Department of Physics, Tsinghua University, Beijing 100084, China}


\begin{abstract}
We propose to use slow muons facilities combined with cyclotron radiation detection for precision test of the weak interaction in the muon decays. Slow positive muon bunches are first injected into a cylindrical superconducting vacuum chamber with uniform strong axial magnetic fields to radially confine the muons. The positrons resulting from muon decays can be detected by their cyclotron radiation, which can be transported to low-noise electronic devices through waveguides coupled to the chamber. The decay positron's energy can be precisely measured down to eV level in the low energy region, which is sensitive to new physics effects such as Majorana neutrinos and new structures of weak interactions.
\end{abstract}

\maketitle


Neutrinos are perhaps the most mysterious particles, due to their weak interactions with other Standard Model (SM) particles. Their small masses are confirmed by neutrino oscillation experiments and direct measurements as well~\cite{PDG2022}, which makes it possible to check their types, because a  Dirac and a Majorana neutrino are fundamentally different if they are massive. The origin of the neutrino mass is one of the unsolved puzzles and indicates some physics beyond the SM such as the seesaw mechanism~\cite{Atre:2009rg}, which explains why the observed neutrino mass is so much smaller in comparison with charged leptons and quarks. On the other hand, it is also not clear why the left-handed $V-A$ interaction is favored in SM. In some extensions of SM~\cite{Pati:1974yy, Mohapatra:1974gc, Senjanovic:1975rk, Mohapatra:1979ia, Mohapatra:1980yp}, the gauge group contains $SU(2)_L \times SU(2)_R \times U(1)$ and the $V+A$ interaction is also allowed. It is due to the breakdown of the left-right symmetry when going from high to low energy regimes that $V-A$ becomes dominant. Various new physics models can be tested with the lepton decay processes. In a generic $V\pm A$ interaction model, the effective Hamiltonian for a lepton decay $\ell^- \to \ell^{\prime -} \bar{\nu}_{\ell'} \nu_l$ with vector currents can be expressed as~\cite{Doi:1981sn, Doi:2005pm}

\begin{equation}
\begin{aligned}
\label{equ:hamiltonian}
\mathcal{H}  = & 4 \frac{G_F}{\sqrt{2}} \left( g_{LL} J^\dagger_{\ell'L\mu} J^\mu_{\ell L} + g_{RR} J^\dagger_{\ell'R\mu} J^\mu_{\ell R} + \right. \\
& \left. g_{LR} J^\dagger_{\ell'L\mu} J^\mu_{\ell R}  + g_{RL} J^\dagger_{\ell'R\mu} J^\mu_{\ell L} \right),
\end{aligned}
\end{equation}
%
where $G_F$ is the Fermi coupling constant, $g_{XX}$ is the relative coupling, $L$($R$) labels the left-handed (right-handed) chirality, and

\begin{equation}
\begin{aligned}
\label{equ:current}
J^{\mu\dagger}_{\ell L} = \sum_j \bar{\ell} \frac{1}{2}\gamma^\mu(1-\gamma^5) U_{\ell j} N_{j L}, \\
J^{\mu\dagger}_{\ell R} = \sum_j \bar{\ell} \frac{1}{2}\gamma^\mu(1+\gamma^5) V_{\ell j} N_{j R}, 
\end{aligned}
\end{equation}
where $N_j$ is the mass eigenstate spinor for the $j$'th neutrino, and $U_{\ell j}$ ($V_{\ell j}$) the left-handed (right-handed) lepton mixing matrix~\cite{Doi:2005pm}. In Eq.~\ref{equ:hamiltonian}, the second term is due to right-handed gauge boson $W_R$, and the last two terms come from the possible mixing between $W_L$ and $W_R$. To test the structure in Eq.~\ref{equ:hamiltonian}, the muon decay can be used as a clean signal since it is free of complications from the strong interaction and hadronization. Surface muons can be created from stopped positive pion decays near the surface of the production target. After passing through stopping materials in the form of thin foils, the kinetic energy of $\mu^+$ can be as low as under $\sim$100 eV~\cite{Morenzoni:1994xf, PSIbeam}. We propose to measure the decay positrons from the slow muons and test the interaction structure, such as Eq.~\ref{equ:hamiltonian}. If the electron's spin is not detected, the differential decay width of a muon in its rest frame is

\begin{equation}
\begin{aligned}
\label{equ:diff_decay_wid}
\frac{d\Gamma}{dx d\cos\theta} = & \frac{m_\mu}{4\pi^3} W_{e\mu}^4 G_F^2 \sqrt{x^2-x_0^2} \left[ F_{IS}(x) \right. \\
& \left. - \mathcal{P}_\mu \cos\theta F_{AS}(x) \right],
\end{aligned}
\end{equation}
%
where $W_{e\mu}=\left(m_\mu^2+m_e^2\right)/2m_\mu$ is the maximum electron energy when the neutrino mass can be neglected, $x=E_e/W_{e\mu}$ is the reduced electron energy, $x_0=m_e/W_{e\mu}$, $\mathcal{P}_\mu$ is the degree of muon polarization, $\theta$ is the angle between the emitted electron and the muon polarization vector, $F_{IS}$ ($F_{AS}$) is the isotropic (anisotropic) part of the spectrum. If the light neutrino mass is neglected, the isotropic part reads

\begin{equation}
\begin{aligned}
\label{equ:F_IS}
F_{IS}(x) = & I x(1-x) + \frac{2}{9} \rho \left(4x^2-3x-x_0^2\right) \\
& + \eta x_0 (1-x), 
\end{aligned}
\end{equation}
where $\rho$ and $\eta$ are the Michel parameters~\cite{Michel:1949qe, Bouchiat:1957zz}. In the SM, $I=\left|g_{LL}\right|^2=1$, $\rho=3\left|g_{LL}\right|^2/4 = 3/4$, and $\eta=0$. In the general model of Eq.~\ref{equ:hamiltonian}, however, the parameters become~\cite{Marquez:2022bpg}

\begin{equation}
\label{equ:michel}
\begin{aligned}
 I = & \sum_{j,k} \left[ \left| f_{LL}^{jk}\right|^2 + \left| f_{RR}^{jk}\right|^2 + \left| f_{LR}^{jk}\right|^2 + \right. \\
& \left. \left| f_{RL}^{jk}\right|^2 + \epsilon \text{Re} \left( f_{LR}^{jk} f_{LR}^{kj*} + f_{RL}^{jk} f_{RL}^{kj*} \right) \right], \\
\rho = & \sum_{j,k} \frac{3}{4} \left( \left| f_{LL}^{jk}\right|^2 + \left| f_{RR}^{jk}\right|^2 \right), \\
\eta = &  \epsilon \sum_{j,k} \left[ \text{Re} \left( f_{LL}^{jk} f_{RR}^{kj*} \right) \right],
\end{aligned}
\end{equation}
%
where the summation is over indices for which the neutrino mass is light enough for the muon decay to happen, $\epsilon=0$(1) for Dirac (Majorana) neutrinos, and

\begin{equation}
\begin{aligned}
\label{equ:michel2}
 f_{LL}^{jk} = g_{LL} U_{ej}U^*_{\mu k},~~f_{RR}^{jk} = g_{RR} V_{ej}V^*_{\mu k},\\
 f_{LR}^{jk} = g_{LR} U_{ej}V^*_{\mu k},~~f_{RL}^{jk} = g_{RL} V_{ej}U^*_{\mu k}.
\end{aligned}
\end{equation}

Therefore, if the neutrinos are of the Majorana type, the last $\eta$-term in Eq.~\ref{equ:F_IS} receives a non-zero contribution. In addition, if the Majorana neutrino $N_{j}$ is heavy enough, but still kinematically allowed in the muon decay, an extra contribution to the $\eta$-term reads

\begin{equation}
\begin{aligned}
\label{equ:eta_term}
\frac{m_j}{2m_\mu} \left[ x_0(1-x) + x_0\sqrt{1-x_0^2} \right] \sum_{k} \text{Re} \left[ f_{LL}^{jk} f_{RL}^{jk*} \right. \\
\left. + f_{RR}^{jk} f_{LR}^{jk*} + \epsilon \left( f_{LL}^{jk} f_{RL}^{kj*} + f_{RR}^{jk} f_{LR}^{kj*} \right) \right].
\end{aligned}
\end{equation}

Note that if the decay neutrino $N_j$ is heavy, the phase space factor in Eq.~\ref{equ:diff_decay_wid} should be also changed to

\begin{align}
\label{equ:phase_space}
\sqrt{x^2-x_0^2} \left( 1 - \frac{m_j^2}{\left(m_\mu^2 + m_e^2 \right)(1-x)} \right).
\end{align}

\begin{figure*}
\centering
\includegraphics[width=0.4\textwidth]{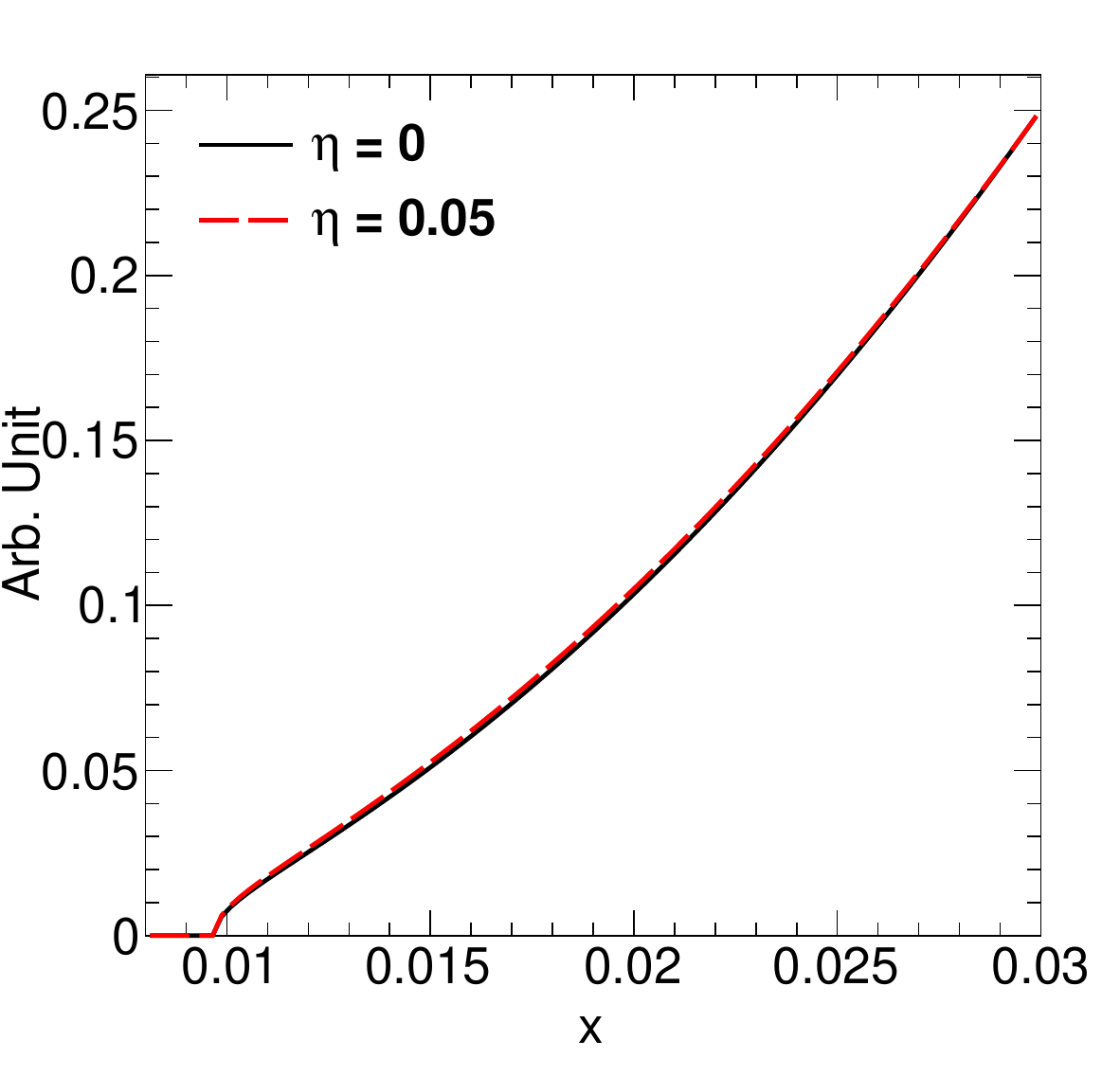}
\includegraphics[width=0.4\textwidth]{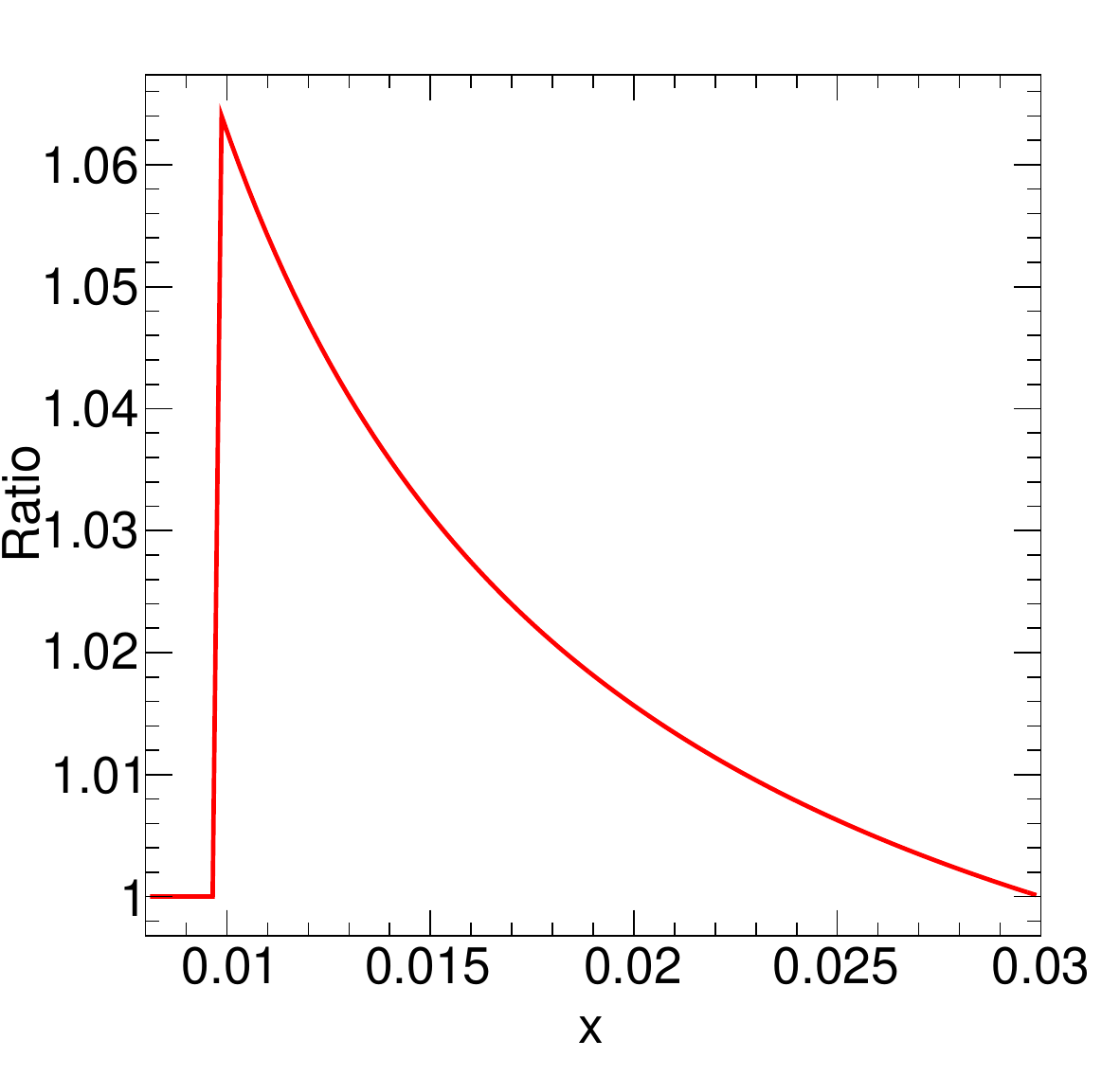}
\caption{Left: The distributions of the reduced energy of the decay electron with $\eta=0$ (solid black) and $\eta=0.05$ (red-dashed). Right: The ratio of the two distributions in the left plot. The red-dashed distribution has been normalized to match the solid black at $x=0.03$. }
\label{fig:eta_comp}
\end{figure*}

Finally, if there are scalar and tensor forms of couplings in Eq.~\ref{equ:hamiltonian}~\cite{Fetscher:1986uj},  extra contributions such as $\frac{1}{2}\text{Re}\left(g_{LL}^V g_{RR}^{S*}\right)$ will appear in $\eta$ irrespective of the neutrino type, where the superscripts $V$ and $S$ represent vector and scalar interactions respectively. The current best measurement for $\eta$ is $-0.0027\pm 0.0050$~\cite{TWIST:2011aa}, which can be improved if low energy decay electrons can be accurately measured, since the effect of the $\eta$-term in Eq.~\ref{equ:F_IS} is most pronounced in the region where $x\to x_0$. For demonstration, the distributions of the reduced energy of the decay electron with $\eta=0$ as in SM and $\eta=0.05$ are shown in Figure~\ref{fig:eta_comp}. The biggest discrepancy happens in the very low $x$ region, where a precise measurement is needed.

\begin{figure*}
\centering
\includegraphics[width=0.65\textwidth]{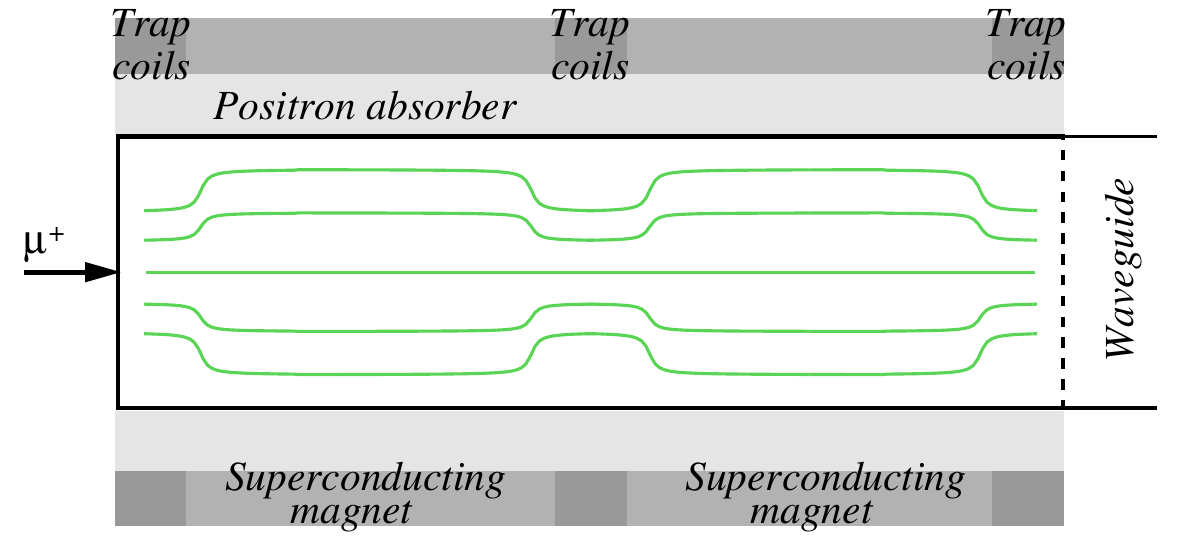}
\caption{The cutaway view of a segment of the proposed experimental apparatus, showing the layer of positron absorber around a cylindrical vacuum chamber, the superconducting magnet, the trap coils, and the waveguide coupled to the end-cap and leading to downstream electronics. The green lines indicate the magnetic field directions.}
\label{fig:apparatus}
\end{figure*}

The measurement of low-$x$ distribution calls for both a high statistics of muon decay events and a precise measurement of the decay electron energy. For the latter, we propose using the cyclotron radiation to measure the electron energy in a non-destructive way~\cite{Monreal:2009za, Project8:2017nal, Project8:2022hun}. The apparatus is illustrated in Figure~\ref{fig:apparatus}. After beam focusing, slow positive muon bunches are injected into a cylindrical vacuum chamber surrounded by a layer of position absorber. The surrounding superconducting magnet provides a 2T uniform magnetic field in the axial direction, which confines the radial motion of muons and decay positrons. To confine the positrons axially, trap coils are periodically placed which render the magnetic field a parabolic dependence of $B=B_0 +\beta (z-z_1)^2 +\beta (z-z_2)^2$ (where $z_{1,2}$ are the longitudinal positions of two adjacent traps) in the axial direction, as indicated by the green lines in Figure~\ref{fig:apparatus}. Due to large Lorentz factors of the decay positrons, not all of them can be axially confined, which leads to some loss of efficiency. However, positrons with a pitch angle $\theta$ (the angle between the positron momentum and the direction of the magnetic field) in a certain range around $\pi/2$ can be confined and detected. 
Conversely, positrons with a small pitch angle will not be confined and will exit the fiducial volume of the chamber quickly. 

The relativistic decay positron will make a helical motion in the chamber. The frequency of its circular cyclotron motion is
\begin{align}
\label{equ:cyclotron_f}
f = \frac{eB}{2\pi m_e \gamma} = \frac{f_0}{\gamma},
\end{align}
where $\gamma$ is the positron Lorentz factor, and $f_0=55.984982$ GHz for $B=2$T.  The decay positron's cyclotron radius can be written as

\begin{align}
\label{equ:cyclotron_R}
R = \frac{m_e c}{eB} \sqrt{\gamma^2-1} \sin\theta.
\end{align}

In a magnetic field of 2T, for a positron with $\theta=\pi/2$ and $E_e = 2$ MeV,  $R=3.23$ mm. Therefore, the inner diameter of the superconducting chamber can be just 1.29 cm if we are only interested in low energy positrons. Any positron with $\sqrt{\gamma^2-1} \sin\theta>3.78$ will hit the inner wall of the chamber. Its shower energy will be dissipated by an absorptive layer surrounding the inner metal wall. The positron's Lorentz factor $\gamma$ can not be very large, since in this case $\sin\theta$ will be small and the positron will be out of the fiducial volume. For positrons with energy below this cut-off, by precisely measuring the cyclotron frequency $f$, the positron energy can be calculated from Eq.~\ref{equ:cyclotron_f}. 


The total power radiated by each positron follows the Liénard extension of the Larmor formula,

\begin{align}
\label{equ:Larmor}
P = \frac{e^4 B^2}{6\pi \epsilon_0 m_e^2 c} (\gamma^2 -1) \sin^2\theta,
\end{align}
where $\epsilon_0$ is the permittivity of free space. The cyclotron frequency of a positron evolving with time can be approximated by

\begin{align}
\label{equ:damp}
f(t) = f(0) \left[ 1 + \frac{\alpha(\gamma^2-1)}{\gamma} t \right], ~\text{with}~\alpha = \frac{e^4 B^2 \sin^2\theta}{6\pi \epsilon_0 m_e^3 c^3},
\end{align}
where $f(0)=f_0/\gamma$ is the positron frequency at the muon decay time. For $\theta=\pi/2$ and $B=2$T, $\alpha=0.770$ s$^{-1}$. This formula is valid provided that $t\ll (\alpha\gamma)^{-1}$. For $\gamma\sim 1$, almost all the radiated power is peaked at the first harmonic of the cyclotron frequency of Eq.~\ref{equ:cyclotron_f}. On the other hand for  $\gamma\gg 1$, most of the power is shifted to the synchrotron radiation with higher frequencies, and the power left for the $n$'th harmonic can be approximated by~\cite{Schwinger:1949ym}

\begin{align}
\label{equ:harmonics}
P_n = \frac{3^{1/6}\Gamma\left(\frac{2}{3}\right)}{4\pi^2} \cdot \frac{e^4 B^2}{\epsilon_0 m_e^2 c \gamma \sqrt{\gamma^2-1}\sin\theta} n^{1/3}.
\end{align}

\begin{figure*}
\centering
\includegraphics[width=0.4\textwidth]{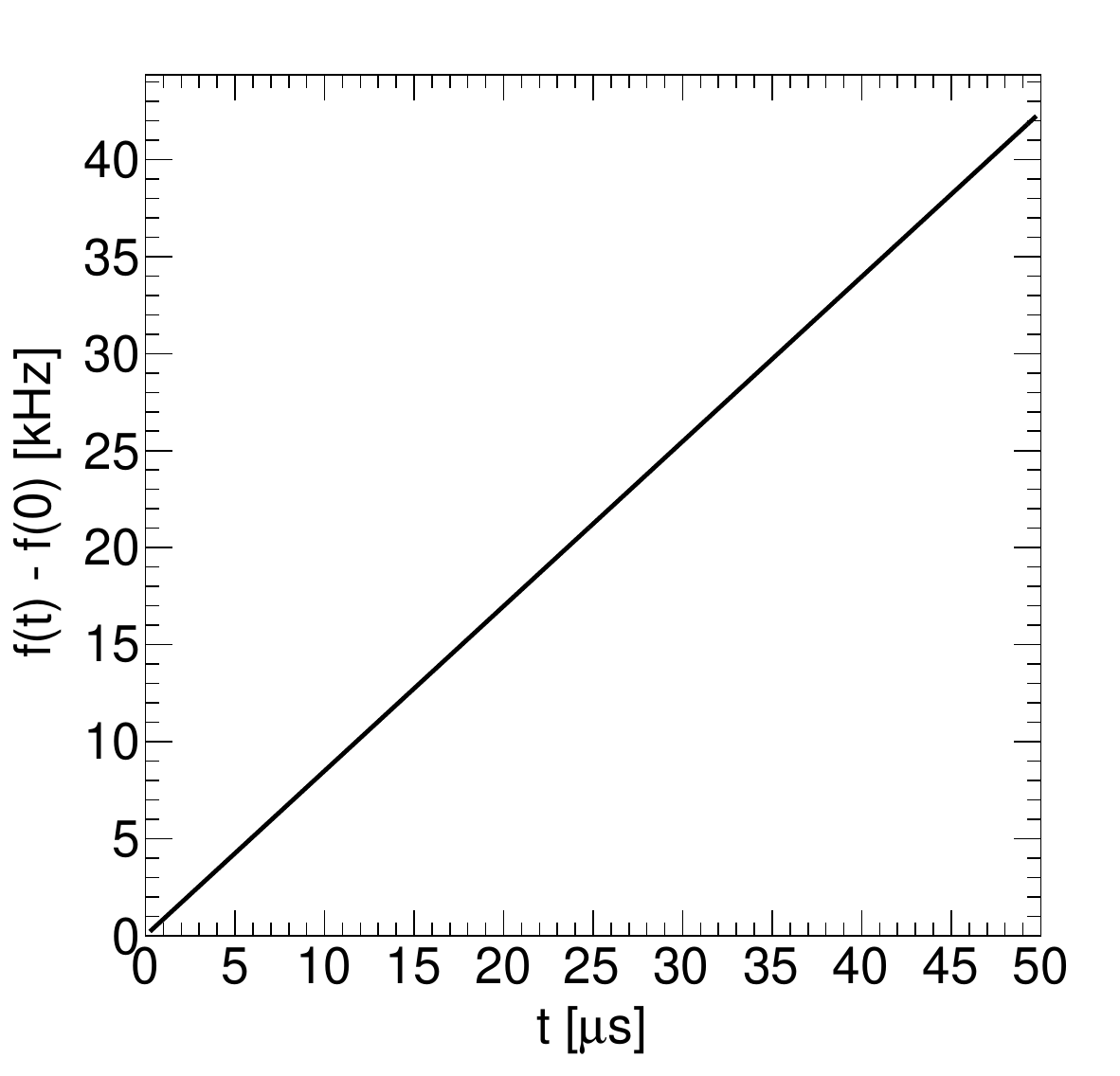}
\includegraphics[width=0.4\textwidth]{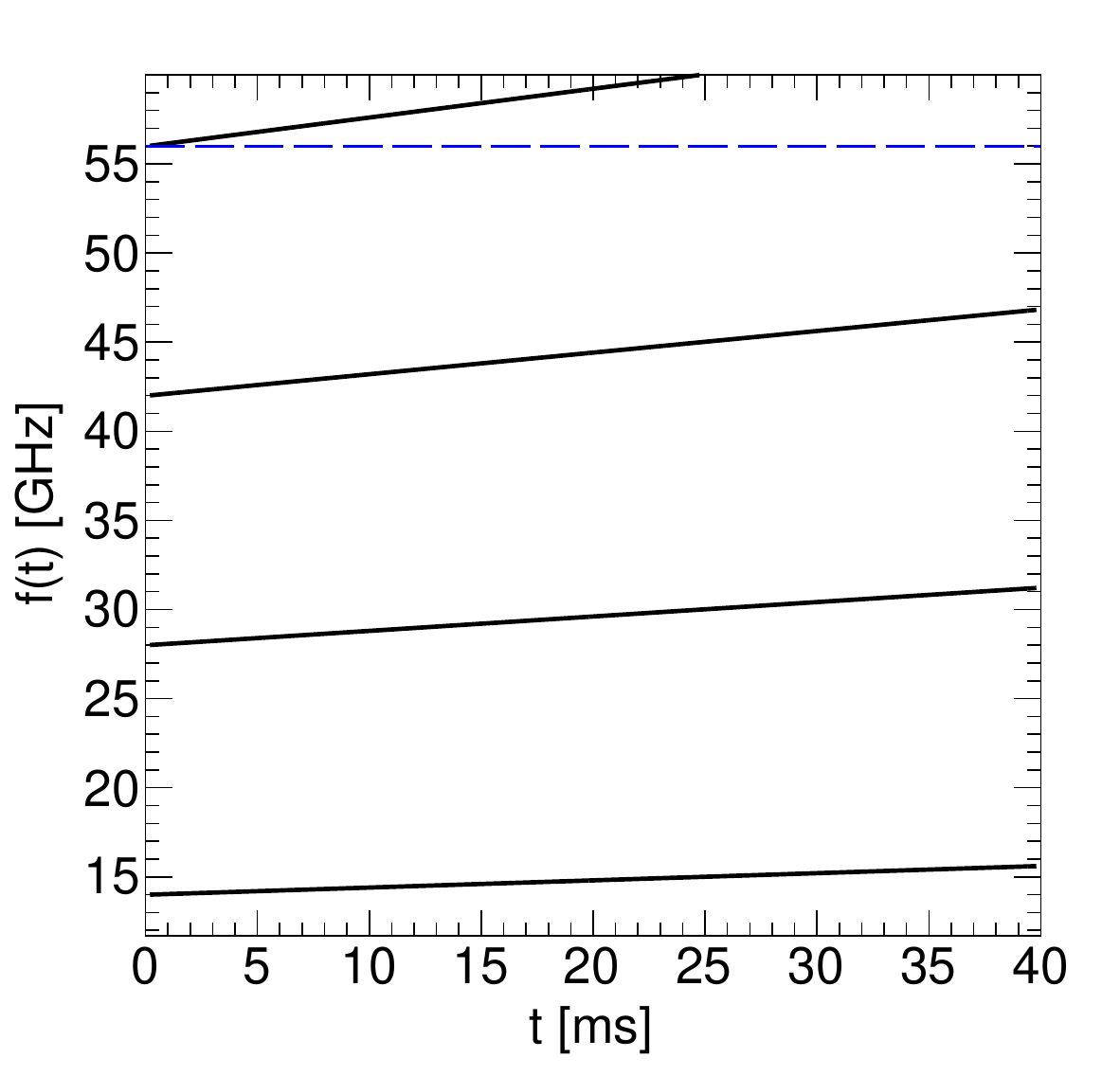}
\caption{The evolution of a decay positron's cyclotron radiation frequency starting from $t=0$ with $B=2$T, $\theta=\pi/2$, $\gamma=1.01$ (left) and $\gamma=4$ (right). For $\gamma=1.01$, only the first harmonic (fundamental) frequency which contains most of the power is shown. For  $\gamma=4$, the first four harmonics are shown, since the positron is relativistic. The blue dashed line indicates the high frequency cut-off at $f_0$.}
\label{fig:frequency}
\end{figure*}

The expected signal frequency of decay positrons with $\theta=\pi/2$ as a function of time is illustrated in Figure~\ref{fig:frequency}. For the case of relativistic positrons, the lowest few harmonics are shown. To remove higher frequencies, a high cut-off at $f_0$ can be applied in the downstream electronics. Any signal with a frequency higher than $f_0$ is rejected, so most of the synchrotron radiation from relativistic positrons can be filtered out. The inner diameter $d$ of the chamber effectively imposes a low frequency cut-off. With $\theta=\pi/2$ and $d=1.29$ cm, positrons with $E_e>2$ MeV (or $f<14.3$ GHz) are rejected. The inner wall of the chamber is also a cylindrical waveguide. The TE$_{11}$ mode of the waveguide also imposes a low cut-off at $1.8412 c/ (\pi d) = 13.6$ GHz. In summary, signals with $14.3~\text{GHz}<f<55.984982~\text{GHz}$ are selected and detected in the end.

For $\theta=\pi/2$ and $\gamma = 1.01$ ($x=0.00977$), the radiated power is $1.27\times 10^{-15}$ W according to Eq.~\ref{equ:Larmor}. For a resolution bandwidth of $\Delta f=200$ kHz and amplifiers with a 20K noise temperature, the thermal noise power is about $5.52\times 10^{-17}$ W according to the formula $kT\Delta f$ ($k$ is the Boltzmann's constant)~\cite{Monreal:2009za}, which is much lower than the signal and the signal detection efficiency would be very high. On the other hand, for $\theta=\pi/2$ and $\gamma = 4$ ($x=0.0387$), the radiated power in the first harmonic frequency is $3.16\times 10^{-15}$ W according to Eq.~\ref{equ:harmonics}, which is also well above the noise level.

\begin{figure*}
\centering
\includegraphics[width=0.4\textwidth]{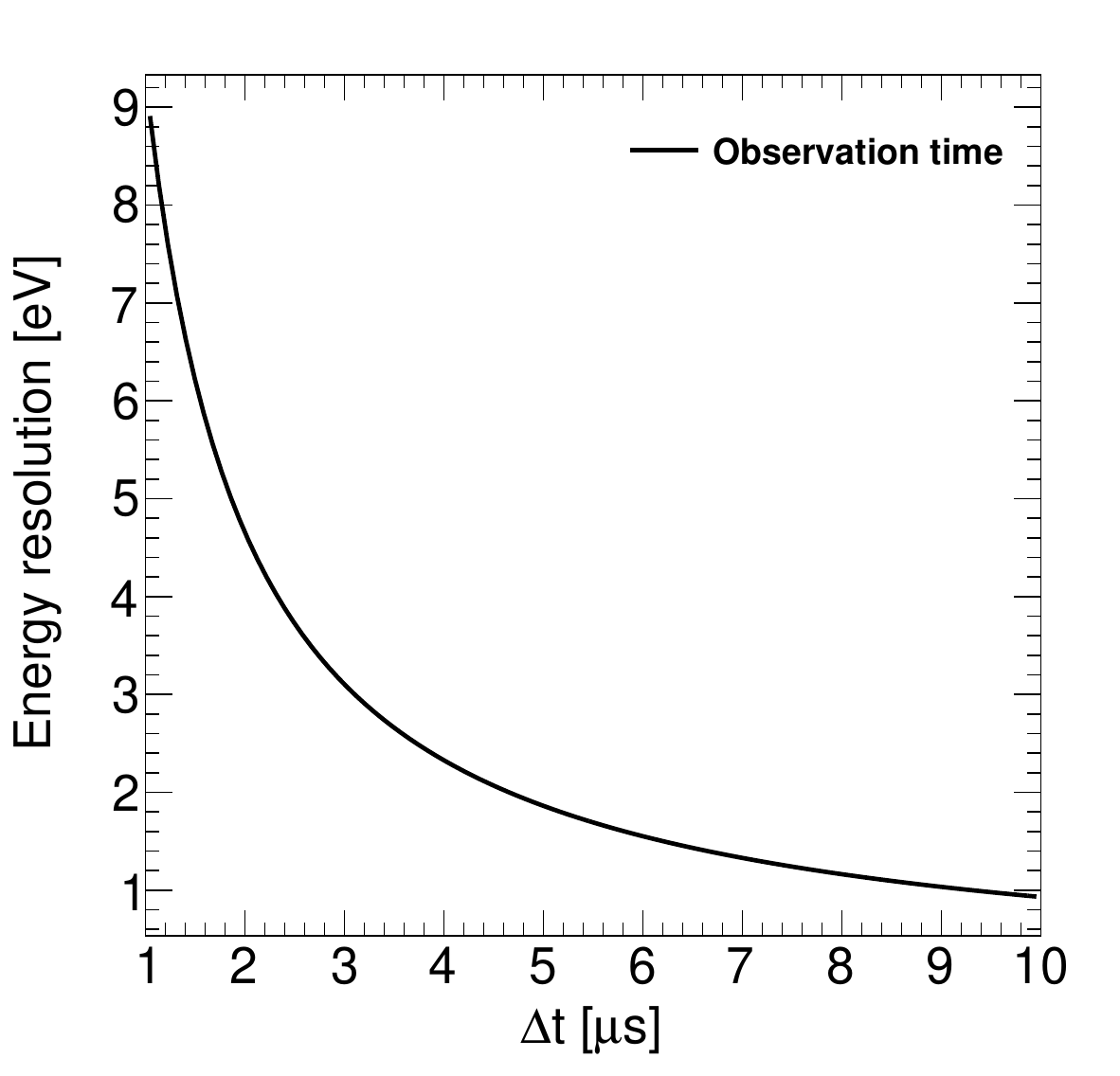}
\includegraphics[width=0.4\textwidth]{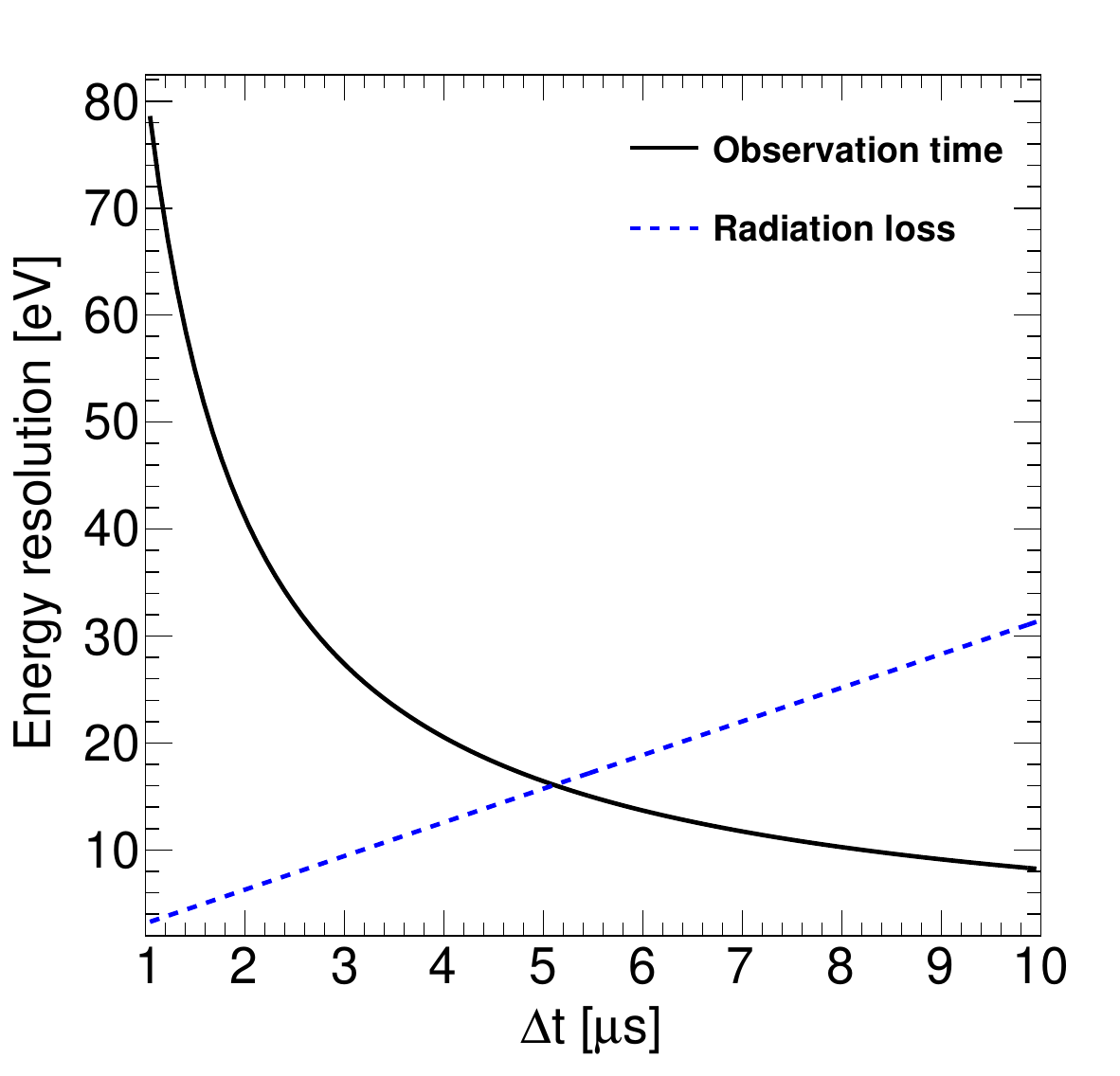}
\caption{The positron energy resolution as a function of the observation time interval $\Delta t$. The black solid (blue dashed) line indicates the resolution due to $\Delta t$ itself (radiation energy loss) for positrons with $\theta=\pi/2$ and $\gamma = 1.01$ (left) and $\gamma = 3$ (right). The resolution due to radiation energy loss is negligible for $\gamma = 1.01$, and therefore it is not shown in the left plot.}
\label{fig:resolution}
\end{figure*}

The energy resolution depends on the observation time interval $\Delta t$ which defines an ``event". There are two effects to be considered. First, $\Delta f\sim 1/\Delta t$ and $\Delta f/f = \Delta E_e/ E_e$~\cite{Project8:2017nal}. Second, the interval $\Delta t$ can not be very large, since the positron loses energy due to radiative emission within $\Delta t$, which also contributes to the energy resolution. The positron energy resolution for $\gamma = 1.01$ and $\gamma = 3$ due to the two effects are shown in Figure~\ref{fig:resolution}. Based on this figure, one can conclude that taking $\Delta t=5$ $\mu$s may be a trade-off among different energy points. For $\gamma\sim 1$, the resolution can reach below 2 eV. It increases to about 42 eV for $\gamma=4$. It is good enough for a precision measurement of the low-$x$ spectrum of the decay positron below $\sim$2 MeV.

We expect the stopped slow muons to have kinetic energy below $\sim$100 eV. A muon with a kinetic energy of 50 eV will travel 0.642 m within its mean lifetime. Therefore, a cylindrical chamber of 1.5 m long will contain about 90\% of the decay positrons. One complication arises due to the axial oscillation of the decay positrons within the fiducial volume, which causes Doppler shifts of the central spectral line and shunts the power to the sideband ones symmetrically distributed around the central one, effectively reducing the signal to background ratio. The relative magnitude of the sideband lines is characterized by the modulation index $h=\Delta\omega/\omega_a$~\cite{AshtariEsfahani:2019yva}, where $\Delta\omega$ is the frequency change due to the Doppler shift and $/\omega_a$ is the axial frequency. When $h\gtrsim 0.5$ a significant power is moved to sidebands. To mitigate this effect, a number of trap coils can be periodically arranged along the length of the chamber so that $\omega_a$ can be increased. Magnetic field inhomogeneities can also broaden the lines and should be minimized during chamber construction.

The magnetic field drifts can be monitored by NMR probes to a precision of $10^{-7}$. To calibrate the apparatus, a low-pressure gaseous $^{83m}Kr$ source can be inserted into the chamber, which provides several narrow electron energy lines with less than 3 eV width, e.g., those at 
30.42 keV, 30.48 keV and 31.94 keV~\cite{Project8:2017nal}. To calibrate the low-noise radiofrequency receiver efficiency, artificial microwave signals of different power can be injected into the waveguide to check the electronic responses. The first event is crucial in determining a positron's energy when its signal first appears in that event. However, as Eq.~\ref{equ:damp} shows, subsequent events can provide the ``slope" of the same signal's energy decay rate with time, therefore provides the value of $\theta$. The detection efficiency as a function of the pitch angle can be quantified thus. It is expected that as the low-$x$ positron loses energy and has its cyclotron radius decreased, its signal strength will fade away, and it will go out of the fiducial volume eventually. Depending on the input muon rate, if needed, to reduce the pileup of relatively high-$x$ positron signals from previous events, applying a voltage pulse periodically to the end-caps of the chamber and resetting the trap coil current simultaneously can flush them away and provide a clean space for subsequent decay events.

\begin{figure*}
\centering
\includegraphics[width=0.45\textwidth]{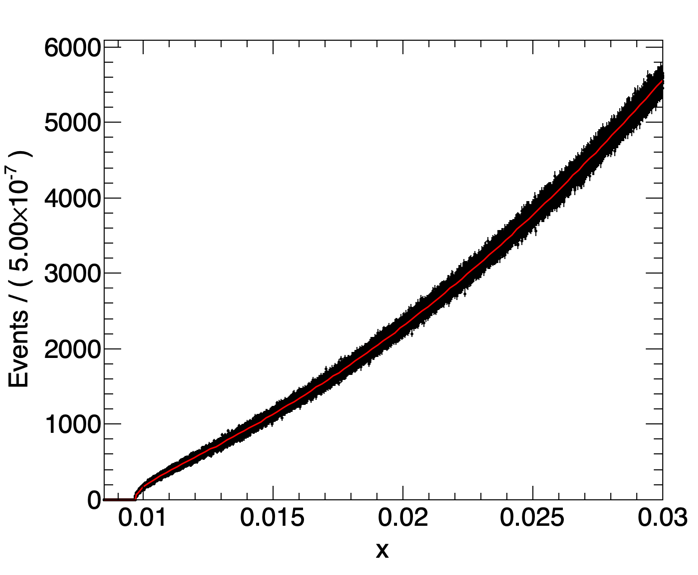}
\caption{The $x$ distribution of the pseudo-data containing $10^6$ events in the range with $x<0.03$ (black markers with error bars), and the fitted function curve (solid red) with the $\eta$ parameter floated. }
\label{fig:fit}
\end{figure*}

If the muon injection rate is $10^5$/s, one expects about 9.6/s decay positrons with $E_e<2$ MeV, and about 4.5/s positrons end up in the region of $x<0.03$, which does not cause a serious event pileup in this scenario. The muon statistics can be much improved if the beam luminosity can be substantially increased~\cite{Mu3e:2020gyw}. As demonstrated in Figure~\ref{fig:fit}, $10^8$ pseudo-data events are generated in $x<0.03$ according to the SM prediction, and a model with the SM plus the $\eta$-term is fitted to this data, which gives $\eta = (6\pm 43)\times 10^{-5}$. The fit uncertainty is improved over the previous result by an order of magnitude, which demonstrates the high sensitivity that the measurement can reach.

In summary, we show that using slow positive muons with kinetic energy $\lesssim 100$ eV, the low part of the decay positron's energy spectrum can be accurately measured by its cyclotron radiation frequency in a constant magnetic field, which is otherwise hard to achieve. The Michel parameter $\eta$ can be measured with a much better precision, which is sensitive to new physics such as Majorana neutrinos and new structures of the weak interactions in the muon decay, if sufficient amount of muon decay events can be collected. 


\begin{acknowledgments}
X. Chen and Z. Hu wish to acknowledge the support from Tsinghua University Initiative Scientific Research Program. We thank Shih-Chieh Hsu from University of Washington for enlightening discussions.
\end{acknowledgments}


\bibliography{muondecay_aps}

\end{document}